\documentclass{nature-Fig}
\usepackage{amsmath}
\usepackage{amssymb}
\usepackage{amsfonts}
\usepackage{graphicx}
\usepackage{pdfpages} 
\usepackage{color}
\usepackage[normalem]{ulem}

\newcommand{\K}[1]{$|#1\rangle$}
\newcommand{\ket}[1]{|#1\rangle}

\newcommand{\qb}[1]{\mbox{qubit #1}}
\newcommand{\bea}{\begin{eqnarray}}
\newcommand{\eea}{\end{eqnarray}}

\newcommand{\s}{{\sigma}}

\title{Tripartite interactions between two phase qubits and a resonant cavity}

\author{F. Altomare$^{1\star}$, J. I. Park$^{1\star}$, K. Cicak$^{1}$, M. A. Sillanp\"a\"a$^{1}\dag$, M. S. Allman$^{1,3}$, D. Li$^{1}$, A. Sirois$^{1,3}$, J.~A.~Strong$^{1,3}$, J. D.
Whittaker$^{1,3}$, and R. W. Simmonds$^{1}$}

\begin{document}
\maketitle
\begin{affiliations}
 \item[$^1$] National Institute of Standards and Technology, 325
Broadway, Boulder CO 80305, USA
 \item[$\dag$] Present address: Helsinki University of Technology, Espoo P.O. Box 2200 FIN-02015 HUT, Finland
 \item[$^3$] University of Colorado, 2000 Colorado Ave, Boulder, CO 80309-0390, USA
 \item[$^\star$] These authors contributed equally to this work.
\end{affiliations}

\begin{abstract}
The creation and manipulation of multipartite entangled states is
important for advancements in quantum computation\cite{Chuang_99}
and communication\cite{Bourennane_98,Zoller_98,Berthiaume_99}, and
for testing our fundamental understanding of quantum
mechanics\cite{GHZ_93} and precision
measurements\cite{Maccone_04}. Multipartite entanglement has been
achieved by use of various forms of quantum bits (qubits), such as
trapped ions\cite{Wineland_05,Blatt_04}, photons\cite{Pan_00}, and
atoms passing through microwave cavities\cite{Haroche_00}. Quantum
systems based on superconducting circuits have been used to
control pair-wise interactions of qubits, either
directly\cite{Nakamura_03,McDermott_05,Steffen_06}, through a
quantum bus\cite{Sillanpaa_07,DiCarlo_09}, or via controllable
coupling\cite{Niskanen_07}. Here, we describe the first
demonstration of coherent interactions of three directly coupled
superconducting quantum systems, two phase qubits and a resonant
cavity. We introduce a simple Bloch-sphere-like representation to
help one visualize the unitary evolution of this tripartite system
as it shares a single microwave photon. With careful control and
timing of the initial conditions, this leads to a protocol for
creating a rich variety of entangled states. Experimentally, we
provide evidence for the deterministic evolution from a simple
product state, through a tripartite W-state, into a bipartite
Bell-state. These experiments are another step towards
deterministically generating multipartite entanglement in
superconducting systems with more than two qubits.
\end{abstract}

\noindent
 With the development of quantum information
science\cite{Chuang_99}, entanglement of multi-particle systems
has become a resource for a new information technology. In
particular, three-particle or tripartite entanglement allows for
teleportation\cite{Bourennane_98}, secret
sharing\cite{Berthiaume_99}, and dense coding\cite{Guo_01}, with
connections to cosmology\cite{Rubens_08}.  Over the last decade,
the development of exquisite control over quantum systems has led
to various demonstrations of tripartite
entanglement\cite{Haroche_00,Pan_00,Blatt_04}. Genuine tripartite
entanglement is delineated by two inequivalent classes of
states\cite{Cirac_00}: GHZ (Greenberger-Horne-Zeilinger) and W,
where the W-state involves only a single photon shared amongst
three systems. Utilizing multipartite entanglement in a
solid-state-qubit system has only recently received theoretical
attention\cite{Nori_06,Martinis_PRA_08,Cho_08}. Thus far in
superconducting systems, bipartite entanglement has been verified
by two-qubit quantum state tomography\cite{Steffen_06} and used to
perform a quantum algorithm\cite{DiCarlo_09}. Spectroscopic
evidence for three-particle entanglement was observed for two
current-biased phase qubits coupled to a lumped element LC-cavity
as well as for Transmon
qubits.\cite{WellstoodPRL05,Wallraff_3qb_08} In the experiments
described below, we first verify the spectroscopic signature of
three coupled systems. Next, we demonstrate coherent interactions.
Frequency detuning of the third system is used to verify the
proper change in the time evolution of two versus three coupled
systems. Finally, we describe a free-evolution process and a
visualization technique as a method for deterministically
preparing arbitrary single-photon tripartite entangled states. We
present evidence for the proper operation of this protocol by
measuring the time-dependent behavior of the two phase qubits.
Here, entanglement is not verified directly, but the data are
consistent with theoretical predictions. Proper execution of this
protocol can prepare the system in a Bell or W-state, as well as
arbitrary entangled states.

\begin{figure}[!p] 
\linespread{1.5}
\center
\includegraphics[trim=30 150 30 150, clip, width=160mm]{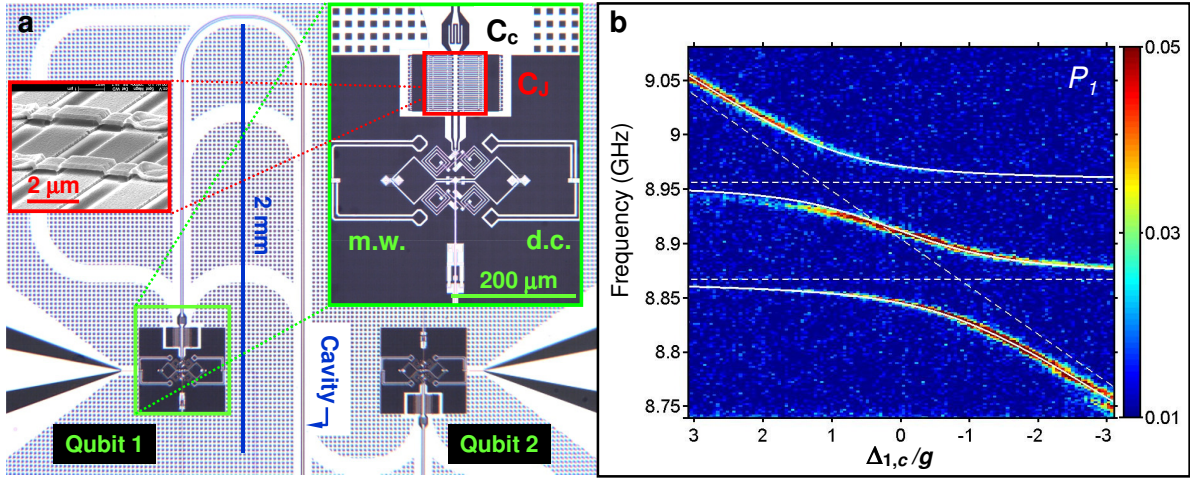}
\caption{Circuit and spectroscopy. \textbf{a}, Optical micrograph
of the electrical circuit with two Josephson phase qubits (\qb{1}
inset overlay right), each with loop inductance $\sim$700 pH and
critical current $\sim 0.91\,\,\mu$A (junction areas \mbox{$\sim 6$
$\mu$m$^2$}) shunted by use of interdigitated capacitors ($C_J\sim
0.7$ pF, including junction capacitance) with vacuum gap
crossovers (inset overlay left), capacitively coupled ($C_c \sim
6.2$ fF) to a coplanar waveguide resonant cavity (of full length
\mbox{$\sim 7$ mm}). The device was fabricated with standard
optical lithography with \mbox{Al/AlO$_\mathrm{x}$/Al} junctions
on a sapphire substrate, with SiO$_2$ as an insulator surrounding
the junctions. \textbf{b}, Microwave spectroscopy of \qb{1} as a
function of detuning $\Delta_{1,c} = \omega_1 - \omega_c$ with
$\omega_2 = \omega_c$. $\Delta_{1,c}$ is varied through the d.c.
flux bias coils and \qb{1} is excited by microwaves applied
through the m.w. (microwave) coil (seen in \textbf{a}). The
intensity color scale represents the probability of \qb{1}
tunneling after the measure pulse. The dashed diagonal line shows
the bare \qb{1} transition frequency. The dashed horizontal lines
represent the two maximally entangled Bell states between \qb{2}
and the cavity.}
\label{Fig_1} 
\end{figure}

 \noindent
 In \mbox{Fig.~\ref{Fig_1}a}, we show an optical micrograph of two
qubits, \qb{1} and \qb{2}, capacitively coupled to either end of
an open-ended coplanar waveguide cavity whose half-wave resonant
mode frequency is $\omega_c/2\pi\approx 8.9$ GHz. These cavities
have shown coherent properties at the single-photon
level.\cite{Sillanpaa_07} Flux-biased phase
qubits\cite{Simmonds_04} can be thought of as anharmonic
LC-oscillators in which a single Josephson junction provides
enough nonlinearity to address the two lowest oscillatory phase
states \K{g} and \K{e}. The energy level separation
$\hbar\omega_j\equiv E_e - E_g$ can be independently tuned over a
range \mbox{$\sim 7$ GHz$-$10 GHz} on the $j$-th qubit by use of
inductively coupled flux bias coils. An additional coil allows us
to apply microwave pulses and fast bias shifts, also used for
single-shot state measurement.\cite{Sillanpaa_07} Independent
state readout on the $j$-th qubit is accomplished by use of an
inductively coupled dc superconducting quantum interference device
(SQUID). We describe this system using a two-qubit Jaynes-Cummings
or Tavis-Cummings model.\cite{TavisCummings_68} In a frame
rotating at an reference frequency $\omega_\mu$, we approximate
the Hamiltonian of the system as \bea H=\hbar\Delta_c a^\dag a +
\sum_{j=1,2} \big[ \hbar\Delta_j \s_j^+ \s_j^- + i \hbar g_j
(\s_j^+ a - a^\dag \s_j^-  ) \big], \eea

\noindent where the mode operators $\s_j^\pm$ and $a^{(\dag)}$
refer to the qubits and the cavity, respectively, with
corresponding detunings $\Delta_j/\hbar\equiv\omega_j-\omega_\mu$
and $\Delta_c/\hbar\equiv\omega_c-\omega_\mu$. Capacitive coupling
$C_c$ between the qubits and the cavity results in an effective
coupling frequency of $2g_j/2\pi\approx\omega_c/2\pi\;
C_c/\sqrt{CC_{\rm J}}\approx 2g/2\pi\sim 90$ MHz for both qubits.
The system exhibits decay rates of \mbox{$\gamma_{1}/2\pi\sim 7$
MHz}, \mbox{$\gamma_{2}/2\pi\sim 10$ MHz}, and
\mbox{$\kappa/2\pi\sim 1$ MHz} for each qubit and the cavity,
respectively. We denote the product of two qubit-cavity states as
$\ket{\eta\eta'n}\equiv\ket{\eta}_1\otimes\ket{\eta'}_2\otimes\ket{n}_c$,
where $\ket{\eta}_j$ label the $j$-th qubit state (\K{g} or \K{e})
and $n$ labels the cavity Fock state.

 \noindent
 The first signature of tripartite interactions is revealed by
spectroscopic measurements\cite{WellstoodPRL05,Wallraff_3qb_08} as
a function of the detuning $\Delta_{1,c}/\hbar=\omega_1-\omega_c$
of \qb{1} when \qb{2} and the cavity are resonant
($\omega_2=\omega_c$). In the case of a single qubit-cavity
system, the Jaynes-Cummings model predicts a single vacuum
Rabi-mode splitting of the qubit state.\cite{Sillanpaa_07} Here,
the single qubit states are split twice by the mutual interaction
of all three systems, as shown in Fig.~\ref{Fig_1}b. We can
interpret this as due to the coupling between the bare \qb{1} and
the antisymmetric pair of maximally entangled Bell states between
\qb{2} and the cavity. The two avoided crossings in the spectrum
occur along the \qb{1} detuning curve, symmetrically displaced
about the tripartite resonance ($\omega_1=\omega_2=\omega_c$).
These measured curves agree well with a full analysis of the
two-qubit Jaynes-Cummings or Tavis-Cummings\cite{Wallraff_3qb_08}
model.

\begin{figure}[!hp] 
\linespread{1.5}
\center
\includegraphics[trim=30 70 30 70, clip,width=160mm]{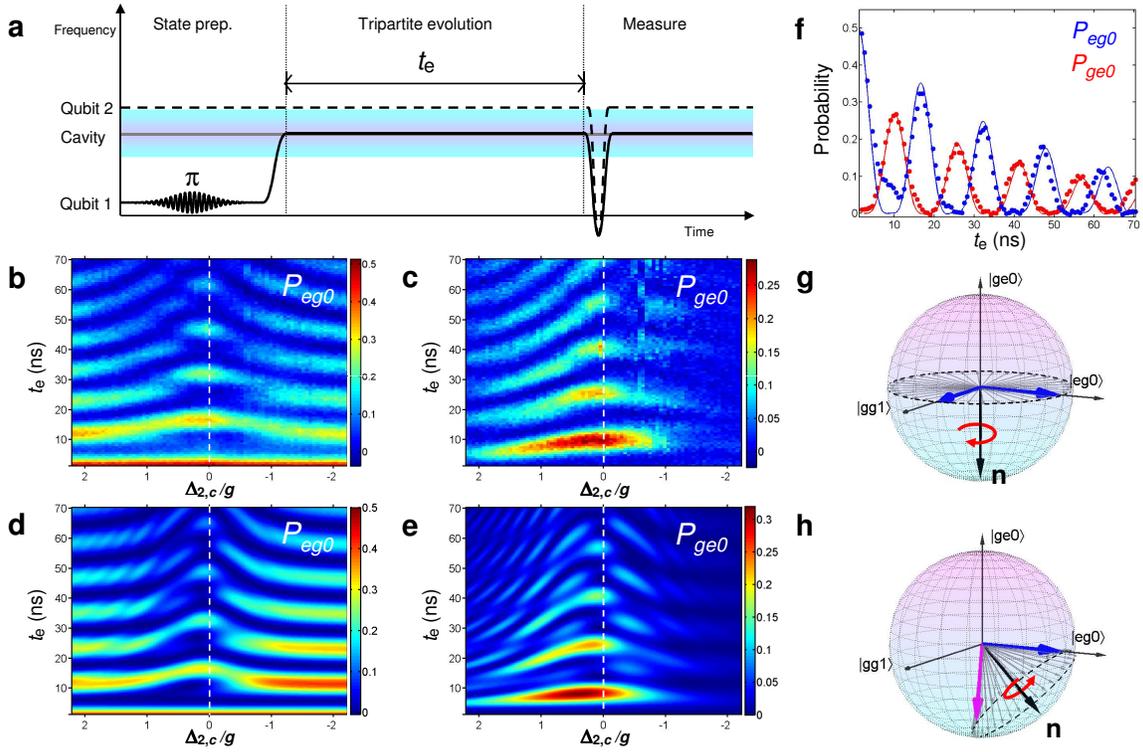}
\caption{Demonstration of basic tripartite interactions.
\textbf{a}, Description for creating a photon in \qb{1} by use of
a $\pi$ pulse, then shifting (solid line) onto resonance with the
cavity  and \qb{2} for various \qb{2} detunings (dashed line).
After an evolution time period $t_{\rm e}$ the qubits are measured
simultaneously\cite{McDermott_05}. \textbf{b},\textbf{c}, Measured
excited state joint probabilities $P_{eg0}$ and $P_{ge0}$ for
states \K{eg0} and \K{ge0}, respectively, during tripartite
interactions after \qb{1} has been excited by a $\pi$ pulse and
shifted onto resonance with the cavity as a function of the
detuning $\Delta_{2,c}=\omega_2-\omega_c$ of \qb{2}.
\textbf{d},\textbf{e}, Theoretical predictions including qubit
visibility and finite rise-time of the shift pulse. The resulting
asymmetry is relatively insensitive to the exact shape of the
shift pulse and can be attributed to additional interference due
to finite detuning of $\sim 0.3g$. Here, we used an exponential
rise-time of $\sim 10$ ns. \textbf{f}, Line cut of the on
resonance tripartite interactions with corresponding theoretical
prediction (solid line). \textbf{g},\textbf{h}, The red arrow
provides a visual cue to the circular trajectory of the tripartite
vector. \textbf{g}, Tripartite sphere representation during simple
vacuum Rabi oscillations of \qb{1}. \textbf{h}, Tripartite sphere
representation during the tripartite evolution from the initial
state \K{eg0}.} \label{Fig_2}
\end{figure}

 \noindent
 With independent control over both qubits, we can easily explore a
convenient state-space whereby a single photon of energy
$\hbar\omega_c$ is shared by our tripartite system. Using a
similar technique established for inducing coherent interactions
between a single qubit and a cavity\cite{Sillanpaa_07}, we
investigate the evolution of vacuum Rabi oscillations between
\qb{1} and the cavity as a function of the detuning of \qb{2}
$\Delta_{2,c}/\hbar=\omega_2-\omega_c$ from the joint
\qb{1}-cavity system ($\omega_1=\omega_c$). For simplicity, we use
the term ``photon" even when describing a single excitation in the
qubit. We begin with both qubits in their ground state and \qb{1}
far off-resonance from the cavity (see pulse diagram in
Fig.\ref{Fig_2}a), then we excite \qb{1} with a photon using a
$\pi$ pulse and bring it onto resonance with the cavity (using a
shift pulse) for a given evolution time period $t_{\rm e}$
followed by simultaneous measurement of both
qubits\cite{McDermott_05}. When \qb{2} is far enough off-resonant,
the resultant vacuum Rabi oscillations are characterized by the
frequency $\Omega_0\equiv 2\,g$, as seen on either side of
\mbox{Fig.~\ref{Fig_2}b,d}. Here, the exchange between \qb{2} and
the \qb{1}-cavity system is energetically prohibited, so that
\qb{1} undergoes basic vacuum Rabi oscillations with the cavity
alone. However, when all three systems are on-resonance with each
other, the photon begins in \qb{1} and then `spreads out' to the
cavity until becoming shared also with \qb{2}, eventually moving
completely to \qb{2}. As time progresses, the photon eventually
returns to \qb{1}.

\noindent
 In this anti-symmetric mode, the oscillation frequency
is given by $\Omega_{\rm a}=\Omega_0/\sqrt{2}$. As the system
evolves, the photon is never completely transferred to the cavity.
There are times when the photon is entirely in \qb{1} or entirely
in \qb{2}, otherwise the system occupies a continuum of entangled
states of both qubits and the cavity (see
\mbox{Fig.~\ref{Fig_2}f}). By measuring the two qubit
simultaneously\cite{McDermott_05}, we can extract the joint
probabilities $P_{eg0}$ and $P_{ge0}$ for single-photon states
\K{eg0} and \K{ge0}, respectively. A theoretical model including
the finite rise-time of the shift pulse ($\sim 10$ ns) agrees well
with the experimental data (\mbox{Fig.~\ref{Fig_2}d-f}).

\noindent
 The above experiment lends itself to a simple geometric
description that can help us visualize the system dynamics. By use
of equation~(1), we can identify the unitary evolution $U(t)=e^{-i
H t/\hbar}$ of the system with a three-dimensional rotation
$R_{\bf n}(\varphi)=e^{-i {\bf n}\cdot {\bf X}\varphi }$ about
${\bf n}\equiv (0,g_2,-g_1)/\sqrt{g_1^2+g_2^2}$ with
$\varphi=\sqrt{g_1^2+g_2^2}\,t$ and ${\bf X}\equiv (X_1,X_2,X_3)$.
Here, $(X_k)_{ij}=-i\epsilon_{ijk}$ helps generate the rotation,
and $\epsilon_{ijk}$ is the totally antisymmetric Levi-Civita
tensor. Time evolution of the system then corresponds to orbits on
a unit sphere azimuthal to the vector ${\bf n}$, where
$(x,y,z)\Leftrightarrow(\ket{gg1},\ket{eg0},\ket{ge0})$, as shown
in \mbox{Fig.~\ref{Fig_2}g,h}. By taking the amplitudes of the
three coupled states as real, absorbing any overall phase into a
redefinition of the states, we can construct a (unit) state vector
analogous to that used for a single spin-1/2 system on the Bloch
sphere. In this case, as the state vector precesses about ${\bf
n}$ and away from any of the coordinate axes, entanglement evolves
over time between all three systems. For the experiment described
above, we start with an initial condition corresponding to the
state \K{eg0}. When \qb{2} is far off-resonance
(\mbox{Fig.~\ref{Fig_2}g}), the system precesses at $\Omega_o$
about ${\bf n}=(0,0,1)$, showing simple vacuum Rabi oscillations
between \qb{1} and the cavity involving the states \K{eg0} and
\K{gg1}, generating bipartite entanglement. However, when all
three systems are on-resonance ($\omega_1=\omega_2=\omega_c$),
$H=g_2 X_2 - g_1 X_3=g(X_2-X_3)$, ${\bf n}=(0,1,-1)/\sqrt{2}$, and
$\varphi=\sqrt{2}g\,t$ leading to a ``tripartite evolution". Now
the initial state vector \K{eg0} precesses about ${\bf n}$ so that
the trajectory passes from the \K{eg0}-axis into a region where
the photon is shared with the cavity and then through the
\mbox{$-$\K{ge0}-axis} (see \mbox{Fig.~\ref{Fig_2}h}). The
oscillations in the two qubits then follow the anti-symmetric mode
frequency $\Omega_{\rm a}$. This single-photon ``tripartite
sphere" representation provides an intuitive picture for
visualizing the equivalence between entangled states in the same
class.\cite{Cirac_00} In this case, the local operations are
vacuum Rabi oscillations or tripartite evolution. We can see that
any arbitrary single-photon tripartite state can be created and
subsequently transformed into any other state on the tripartite
sphere, much like unitary operations and rotations on the Bloch
sphere. Of particular interest is the fact that a specific initial
state will follow a specific trajectory under tripartite
evolution, transforming the amount of entanglement continuously.
Below, we determine the conditions for directly demonstrating
transformations between Bell and W states, starting from an
initially pure state.

\begin{figure}[!p] 
\center
\includegraphics[trim=40 150 40 150, clip,width=160mm]{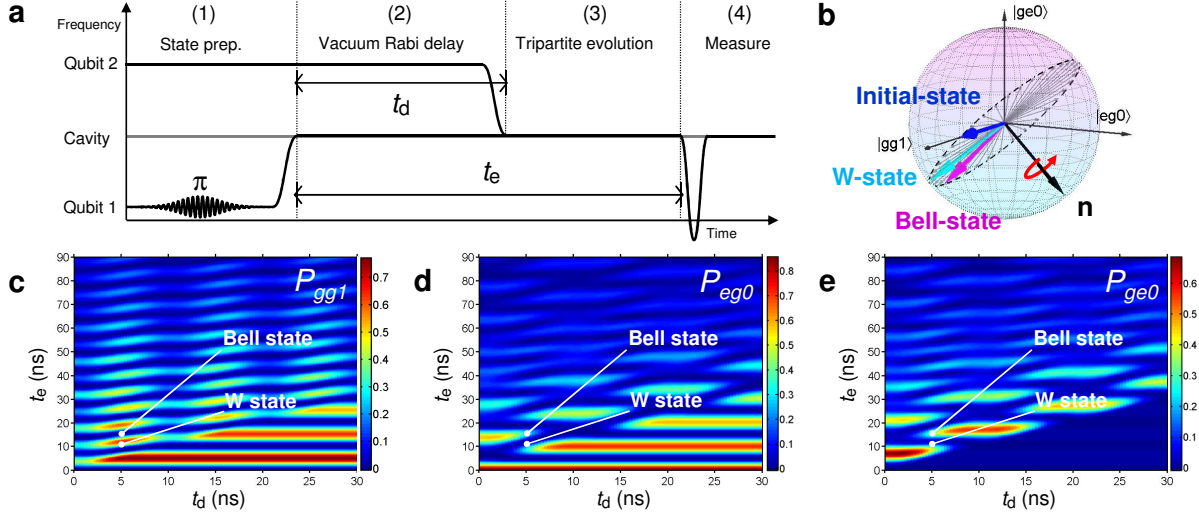}
\caption{Experimental protocol and theoretical predictions for
generating generalized arbitrary single photon tripartite
evolutions. \textbf{a}, Pulse sequence: (1) A photon is inserted
in the system by exciting \qb{1}. (2) A shift pulse brings \qb{1}
onto resonance with the cavity, producing vacuum Rabi
oscillations. (3) A shift pulse brings \qb{2} onto resonance after
the delay time $t_{\rm d}$, initiating tripartite interactions
that evolve over a time period $t_{\rm e}-t_{\rm d}$. (4) Both
qubits are measured simultaneously. \textbf{b}, Tripartite sphere
representation of the tripartite evolution for the initial state
\K{gg1} prepared during a delay time period $t_{\rm
d}=\pi/\Omega_o$. The red arrow provides a visual cue to the
circular trajectory of the tripartite vector. \textbf{c},
Predicted state occupation of one photon in the resonant cavity.
\textbf{d},\textbf{e}, Predicted joint state probabilities
$P_{eg0}$ and $P_{ge0}$ for measurement of \qb{1} and 2 as
functions of both $t_{\rm d}$ and $t_{\rm e}$. Blue color
represents low values, red represents high. The simulations were
performed with a finite energy relaxation time, a finite shift
pulse rise-time ($\sim 10$ ns), and some residual qubit detuning
of $\sim 0.3g$ from the cavity.} \label{Fig_3}
\end{figure}

 \noindent
 We begin with a single photon in \qb{1} or \qb{2}. As
shown above, vacuum Rabi oscillations represent arbitrary
rotations in the \K{gg1}-\K{eg0} plane (between the cavity and
\qb{1}) or the \K{gg1}-\K{ge0} plane (cavity and \qb{2}). These
two operations in succession allow us complete access to the
\K{eg0}-\K{ge0} plane, and, thus, the ability to prepare any
initial state on the entire single-photon tripartite sphere. In
order to generate Bell and W states, we can start with the photon
in the cavity, \K{gg1}. Under tripartite evolution the system
passes first through the W state, $\ket{\rm
W}\equiv(-\ket{gg1}+\ket{eg0}+\ket{ge0})/\sqrt{3}$, and then
through the Bell state, $\ket{\rm Bell}\equiv-
(\ket{eg0}+\ket{ge0})/\sqrt{2}$, as the system vector rotates
about the n-vector, ${\bf n}=(0,1,-1)/\sqrt{2}$ as shown in
Fig.~\ref{Fig_3}b. In total, the system will pass through two Bell
states and four W states for one full revolution about ${\bf n}$.
The frequency $\Omega_{\rm s}=\sqrt{2}\,\Omega_0$ of qubit
oscillations follows from the definition of $\varphi$ and the arc
traced out by the system trajectory. In this symmetric mode
($\Omega_{\rm s}=2\Omega_{\rm a}$) the photon ``splits" as it
leaves the cavity, having an equal probability for going to \qb{1}
or \qb{2}, and subsequently returning completely to the cavity.

\noindent
 Experimentally, we sample a variety of initial states
by allowing \qb{1} (which initially has the photon) to undergo
vacuum Rabi oscillations with the cavity for a delay time period
$t_{\rm d}$ before we bring \qb{2} into tripartite resonance.
\mbox{Fig.~\ref{Fig_3}c-e} shows a prediction for the unitary
evolution of the system for nearly a continuum of values for
$t_{\rm d}$. Here, the joint probabilities are $P_{gg1}$,
$P_{eg0}$, and $P_{ge0}$ for states \K{gg1}, \K{eg0}, and \K{ge0},
respectively. Notice that for $t_{\rm d} = 2\pi/\Omega_0$, the
system will exhibit the anti-symmetric mode (indicated along the
dashed line) as described earlier. However when $t_{\rm
d}=\pi/\Omega_0$, we prepare the (initial condition) \K{gg1},
allowing for a tripartite evolution of the symmetric mode. After a
period of time $t_e=\pi/4\Omega_{\rm s}$, the excited-state
probability for both qubits is 1/3 and the system is in the \K{\rm
W}-state, with the photon equally distributed among the two qubits
and cavity. After a period of time $t_e=\pi/2\Omega_{\rm s}$, the
excited-state probability for both qubits is 1/2 and the system is
in the bipartite \K{\rm Bell}-state. These points are indicated in
\mbox{Fig.~\ref{Fig_3}c-e}, with the first three states shown as
vectors on the tripartite sphere in \mbox{Fig.~\ref{Fig_3}b}. Here
the simulations have included finite energy relaxation and the
rise-time of the shift pulses.

\begin{figure}[!p] 
\center
\includegraphics[trim=50 100 30 100, clip,width=160mm]{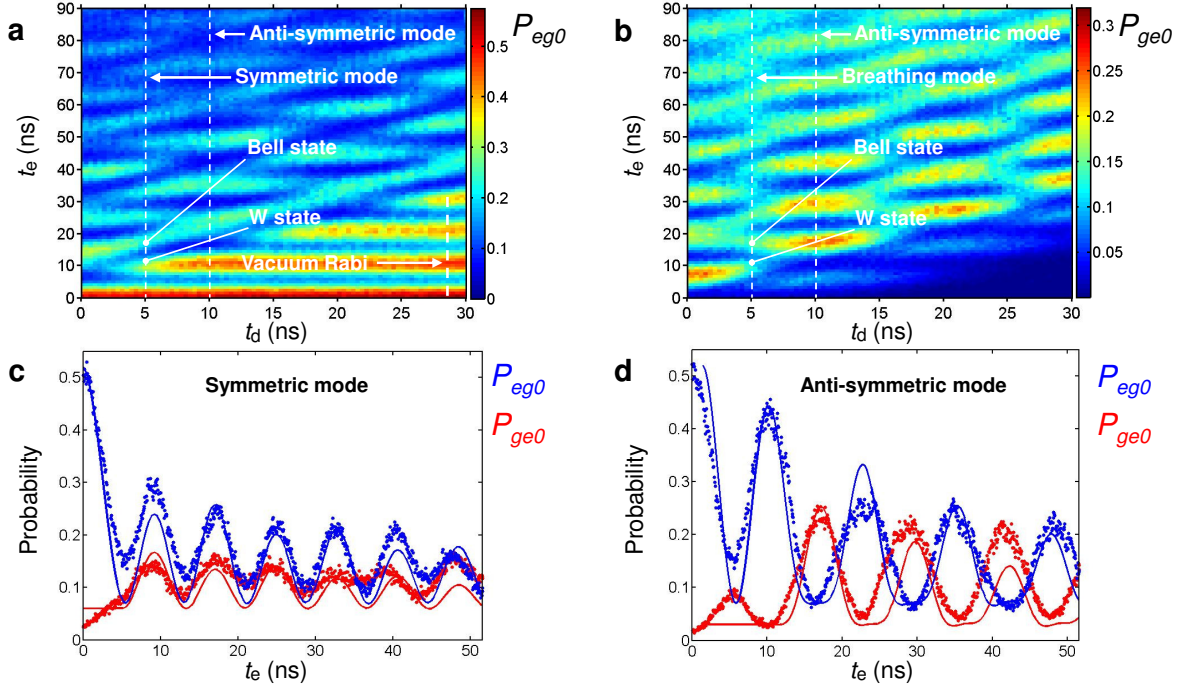}
\caption{Experimental demonstration of arbitrary tripartite
interactions between both phase qubits and the cavity.
\textbf{a},\textbf{b}, Measured joint state probabilities
$P_{eg0}$ and $P_{ge0}$ for measurement of \qb{1} and 2 as
functions of both $t_{\rm d}$ and $t_{\rm e}$. \textbf{c},
Extracted curves (along dashed line) for the initial state \K{gg1}
producing a tripartite evolution of the symmetric mode, showing in
phase oscillations along with theoretical predictions (solid
lines) from \mbox{Fig.~\ref{Fig_3}d,e}. During this evolution, the
system's entanglement continuously transforms starting from a pure
state \K{gg1}. In the ideal case, the system evolves through W and
Bell states. \textbf{d}, Extracted curves (along dashed line) for
the initial state \K{eg0} producing a tripartite evolution of the
anti-symmetric mode, showing out of phase oscillations along with
theoretical predictions (solid lines) from
\mbox{Fig.~\ref{Fig_3}d,e}.} \label{Fig_4}
\end{figure}

\noindent
 We simultaneously measure both qubits and observe the
occupation probabilities of the two qubits over time as they
evolve from a continuum of initial states, superposition states of
\qb{1} and the cavity. Although possible, as explained later, we
do not measure the cavity state directly.
\mbox{Fig.~\ref{Fig_4}c,d} shows extracted line cuts from
\mbox{Fig.~\ref{Fig_4}a,b} for two initial conditions (dashed
lines) corresponding to the symmetric and anti-symmetric modes. As
can been seen from \mbox{Fig.~\ref{Fig_3}d,e}, the theoretical
predictions for the evolutions agree with the measurements. For
the symmetric mode we find in-phase oscillations of the two qubits
at $\Omega_{\rm s}\sim\sqrt{2}\,\Omega_0$, while for the
anti-symmetric mode, we find that the two qubits oscillate out of
phase with each other with the anti-symmetric mode frequency
$\Omega_{\rm a}\sim\Omega_0/\sqrt{2}$, where $\Omega_0$ is the
frequency of the vacuum Rabi oscillations that occur during the
delay time period $t_{\rm d}$ (lower right hand corner of
\mbox{Fig.~\ref{Fig_4}a}). The measured frequencies agree within
$\sim 15\,\%$ of the ideal case, due to the finite rise-time of
the shift pulses and some residual nonzero detuning of each qubit
frequency. A theoretical model including these imperfections
(solid lines) agrees well with the data.

 \noindent
In the present experiment, we improved the previous
design\cite{Sillanpaa_07,QIP_09} by reducing the qubit junction
areas to reduce the number of two-level system defects. This more
than doubled the qubit visibility and provided the necessary
`clean' cavity region for observing tripartite interactions.
However, it was not possible to perform two-qubit state tomography
over the required time scales due to short relaxation
times\cite{MartinisPRL05} matched with the continued presence of
two-level system defects that limited the qubit
visibility\cite{Cooper_04} to $\lesssim 50\,\%$. With further
reductions in junction size, we can raise the single qubit
visibility to 90\% allowing full tomographic characterization of
both qubits.\cite{Steffen_06}

\noindent
 In the future, we intend to perform correlated
measurements and tomography of this tripartite system. This
requires a fast, dispersive measurement and readout of the qubits
to solve three difficulties. First, the tunneling-based
measurement of either qubit will populate the cavity with
unwanted photons due to a crosstalk process.\cite{McDermott_05}
Second, a dispersive measurement will increase qubit visibility
ensuring clear tomography. And third, after measurement of the two
qubits, subsequent qubit rotations will ensure proper state
preparation for one of the qubits, making it ready for
re-interaction with the cavity. In this way, we can reuse one of
the qubits through state transfer\cite{Sillanpaa_07}, to fully
determine the cavity state\cite{Hofheinz_09}. Improvements are
currently underway to modify our slow switching-current SQUID
readout to a fast, dispersive resonant readout.

 \noindent
Tripartite interactions provide a means of engineering qubit
entangled states. We have provided a theoretical description of
the two-qubit Jaynes-Cummings or Tavis-Cummings model and a
free-evolution protocol for the deterministic preparation of Bell,
W, or arbitrary entangled states with one shared photon. This
includes a convenient geometric description helpful for
visualizing the unitary evolution. We have presented clear
experimental evidence of tripartite interactions between two

 superconducting phase qubits and a resonant microwave cavity. Our
system has a fundamental advantage over strictly multi-qubit
systems\cite{Nori_06,Martinis_PRA_08,Cho_08}: The cavity provides
a larger degree of freedom as opposed to the two discrete levels
associated with qubits and can be thought of as a multiphoton
register or entanglement resource. Arbitrary preparation of
multiphoton states in this cavity via one of the
qubits\cite{Hofheinz_09}, and subsequent interactions for
entanglement distribution will allow for the creation of further
classes of entanglement, such as a GHZ state.


\begin{addendum}
\item[Acknowledgements] This work was financially supported by
NIST and DTO, and the Academy of Finland. Contribution of the U.S.
 government, not subject to copyright.
\end{addendum}

\begin{addendum}
\item[Author Information] Reprints and permissions information is
available at www.nature.com/reprints. The authors declare no
competing financial interests. Correspondence and requests for
materials should be addressed to R.W.S.
(simmonds@boulder.nist.gov).
\end{addendum}

\begin{thebibliography}{10} \expandafter\ifx\csname url\endcsname\relax   \def\url#1{\texttt{#1}}\fi \expandafter\ifx\csname urlprefix\endcsname\relax\def\urlprefix{URL }\fi \providecommand{\bibinfo}[2]{#2} \providecommand{\eprint}[2][]{\url{#2}} \bibitem{Chuang_99} \bibinfo{author}{Gottesman, D.} \& \bibinfo{author}{Chuang, I.~L.} \newblock \bibinfo{title}{Demonstrating the viability of universal quantum   computation using teleportation and single-qubit operations}. \newblock \emph{\bibinfo{journal}{Nature}} \textbf{\bibinfo{volume}{402}},   \bibinfo{pages}{390} (\bibinfo{year}{1999}). \bibitem{Bourennane_98} \bibinfo{author}{Karlsson, A.} \& \bibinfo{author}{Bourennane, M.} \newblock \bibinfo{title}{Quantum teleportation using three-particle   entanglement}. \newblock \emph{\bibinfo{journal}{Phys. Rev. A}} \textbf{\bibinfo{volume}{58}},   \bibinfo{pages}{4394} (\bibinfo{year}{1998}). \bibitem{Zoller_98} \bibinfo{author}{Briegel, H.~J.}, \bibinfo{author}{D{\"u}r, W.},   \bibinfo{author}{Cirac, J.} \& \bibinfo{author}{Zoller, P.} \newblock \bibinfo{title}{Quantum repeaters: The role of imperfect local   operations in quantum communication}. \newblock \emph{\bibinfo{journal}{Phys. Rev. Lett.}}   \textbf{\bibinfo{volume}{81}}, \bibinfo{pages}{5932} (\bibinfo{year}{1998}). \bibitem{Berthiaume_99} \bibinfo{author}{Hillery, M.}, \bibinfo{author}{Buzek, V.} \&   \bibinfo{author}{Berthiaume, A.} \newblock \bibinfo{title}{Quantum secret sharing}. \newblock \emph{\bibinfo{journal}{Phys. Rev. A}} \textbf{\bibinfo{volume}{59}},   \bibinfo{pages}{1829} (\bibinfo{year}{1999}). \bibitem{GHZ_93} \bibinfo{author}{Greenberger, D.~M.}, \bibinfo{author}{Horne, M.~A.} \&   \bibinfo{author}{Zeilinger, A.} \newblock \bibinfo{title}{Multiparticle interferometry and the superposition   principle}. \newblock \emph{\bibinfo{journal}{Physics Today}}   \textbf{\bibinfo{volume}{46}}, \bibinfo{pages}{22--29}   (\bibinfo{year}{1993}). \bibitem{Maccone_04} \bibinfo{author}{Giovannetti, V.}, \bibinfo{author}{Lloyd, S.} \&   \bibinfo{author}{Maccone, L.} \newblock \bibinfo{title}{Quantum-enhanced measurements: Beating the standard   quantum limit}. \newblock \emph{\bibinfo{journal}{Science}} \textbf{\bibinfo{volume}{306}},   \bibinfo{pages}{1330} (\bibinfo{year}{2004}). \bibitem{Wineland_05} \bibinfo{author}{Leibfried, D.} \emph{et~al.} \newblock \bibinfo{title}{Creation of a six-atom '{S}chr{\"o}dinger cat'   state}. \newblock \emph{\bibinfo{journal}{Nature}} \textbf{\bibinfo{volume}{438}},   \bibinfo{pages}{639--641} (\bibinfo{year}{2005}). \bibitem{Blatt_04} \bibinfo{author}{Roos, C.~F.} \emph{et~al.} \newblock \bibinfo{title}{Control and measurement of three-qubit entangled   states}. \newblock \emph{\bibinfo{journal}{Science}} \textbf{\bibinfo{volume}{304}},   \bibinfo{pages}{1478--1482} (\bibinfo{year}{2004}). \bibitem{Pan_00} \bibinfo{author}{Pan, J.-W.}, \bibinfo{author}{Bouwmeester, D.},   \bibinfo{author}{Daniell, M.}, \bibinfo{author}{Weinfurter, H.} \&   \bibinfo{author}{Zeilinger, A.} \newblock \bibinfo{title}{Experimental test of quantum nonlocality in   three-photon {G}reenberger-{H}orne-{Z}eilinger entanglement}. \newblock \emph{\bibinfo{journal}{Nature}} \textbf{\bibinfo{volume}{403}},   \bibinfo{pages}{515--519} (\bibinfo{year}{2000}). \bibitem{Haroche_00} \bibinfo{author}{Rauschenbeutel, A.} \emph{et~al.} \newblock \bibinfo{title}{Step-by-step engineered multiparticle entanglement}. \newblock \emph{\bibinfo{journal}{Science}} \textbf{\bibinfo{volume}{288}},   \bibinfo{pages}{2024--2028} (\bibinfo{year}{2000}). \bibitem{Nakamura_03} \bibinfo{author}{Yamamoto, T.}, \bibinfo{author}{\mbox{Yu}. A.~Pashkin},   \bibinfo{author}{Astafiev, O.}, \bibinfo{author}{Nakamura, Y.} \&   \bibinfo{author}{Tsai, J.~S.} \newblock \bibinfo{title}{Quantum oscillations in two coupled charge qubits}. \newblock \emph{\bibinfo{journal}{Nature}} \textbf{\bibinfo{volume}{425}},   \bibinfo{pages}{941} (\bibinfo{year}{2003}). \bibitem{McDermott_05} \bibinfo{author}{McDermott, R.} \emph{et~al.} \newblock \bibinfo{title}{Simultaneous state measurement of coupled {J}osephson   phase qubits}. \newblock \emph{\bibinfo{journal}{Science}} \textbf{\bibinfo{volume}{307}},   \bibinfo{pages}{1299--1302} (\bibinfo{year}{2005}). \bibitem{Steffen_06} \bibinfo{author}{Steffen, M.} \emph{et~al.} \newblock \bibinfo{title}{Measurement of the entanglement of two   superconducting qubits via state tomography}. \newblock \emph{\bibinfo{journal}{Science}} \textbf{\bibinfo{volume}{313}},   \bibinfo{pages}{1423--1425} (\bibinfo{year}{2006}). \bibitem{Sillanpaa_07} \bibinfo{author}{Sillanp{\"a}{\"a}, M.~A.}, \bibinfo{author}{Park, J.~I.} \&   \bibinfo{author}{Simmonds, R.~W.} \newblock \bibinfo{title}{Coherent quantum state storage and transfer between   two phase qubits via a resonant cavity}. \newblock \emph{\bibinfo{journal}{Nature}} \textbf{\bibinfo{volume}{449}},   \bibinfo{pages}{438--442} (\bibinfo{year}{2007}). \bibitem{DiCarlo_09} \bibinfo{author}{DiCarlo, L.} \emph{et~al.} \newblock \bibinfo{title}{Demonstration of two-qubit algorithms with a   superconducting quantum processor}. \newblock \emph{\bibinfo{journal}{Nature}} \textbf{\bibinfo{volume}{460}},   \bibinfo{pages}{240--244} (\bibinfo{year}{2009}). \bibitem{Niskanen_07} \bibinfo{author}{Niskanen, A.~O.} \emph{et~al.} \newblock \bibinfo{title}{Quantum coherent tunable coupling of superconducting   qubits}. \newblock \emph{\bibinfo{journal}{Science}} \textbf{\bibinfo{volume}{316}},   \bibinfo{pages}{723--726} (\bibinfo{year}{2007}). \bibitem{Guo_01} \bibinfo{author}{Hao, C.-F., J.-C.~Li} \& \bibinfo{author}{Guo, G.-C.} \newblock \bibinfo{title}{Controlled dense coding using the   {G}reenberger-{H}orne-{Z}eilinger state}. \newblock \emph{\bibinfo{journal}{Phys. Rev A}} \textbf{\bibinfo{volume}{63}},   \bibinfo{pages}{054301} (\bibinfo{year}{2001}). \bibitem{Rubens_08} \bibinfo{author}{Borsten, L.}, \bibinfo{author}{Dahanayake, D.},   \bibinfo{author}{Duff, M.~J.}, \bibinfo{author}{Ebrahim, H.} \&   \bibinfo{author}{Rubens, W.} \newblock \bibinfo{title}{Wrapped branes as qubits}. \newblock \emph{\bibinfo{journal}{Phys. Rev. Lett.}}   \textbf{\bibinfo{volume}{100}}, \bibinfo{pages}{251602}   (\bibinfo{year}{2008}). \bibitem{Cirac_00} \bibinfo{author}{D{\"u}r, W.}, \bibinfo{author}{Vidal, G.} \&   \bibinfo{author}{Cirac, J.} \newblock \bibinfo{title}{Three qubits can be entangled in two inequivalent   ways}. \newblock \emph{\bibinfo{journal}{Phys. Rev. A}} \textbf{\bibinfo{volume}{62}},   \bibinfo{pages}{062314} (\bibinfo{year}{2000}). \bibitem{Nori_06} \bibinfo{author}{Wei, L.~F.}, \bibinfo{author}{Liu, Y.-x.} \&   \bibinfo{author}{Nori, F.} \newblock \bibinfo{title}{{G}reenberger-{H}orne-{Z}eilinger entanglement in   superconducting circuits}. \newblock \emph{\bibinfo{journal}{Phys. Rev. Lett.}}   \textbf{\bibinfo{volume}{96}}, \bibinfo{pages}{246803}   (\bibinfo{year}{2006}). \bibitem{Martinis_PRA_08} \bibinfo{author}{Galiautdinov, A.} \& \bibinfo{author}{Martinis, J.~M.} \newblock \bibinfo{title}{Maximally entangling tripartite protocols for   {J}osephson phase qubits}. \newblock \emph{\bibinfo{journal}{Phys. Rev. A}} \textbf{\bibinfo{volume}{78}},   \bibinfo{pages}{010305} (\bibinfo{year}{2008}). \bibitem{Cho_08} \bibinfo{author}{Kim, M.~D.} \& \bibinfo{author}{Cho, S.~Y.} \newblock \bibinfo{title}{Macroscopic {G}reenberger-{H}orne-{Z}eilinger and {W}   states in flux qubits}. \newblock \emph{\bibinfo{journal}{Phys. Rev. B}} \textbf{\bibinfo{volume}{77}},   \bibinfo{pages}{100508} (\bibinfo{year}{2008}). \bibitem{WellstoodPRL05} \bibinfo{author}{Xu, H.} \emph{et~al.} \newblock \bibinfo{title}{Spectroscopy of three-particle entanglement in a   macroscopic superconducting circuit}. \newblock \emph{\bibinfo{journal}{Phys. Rev. Lett.}}   \textbf{\bibinfo{volume}{94}}, \bibinfo{pages}{027003}   (\bibinfo{year}{2005}). \bibitem{Wallraff_3qb_08} \bibinfo{author}{Fink, J.~M.} \emph{et~al.} \newblock \bibinfo{title}{Dressed collective qubit states and the   tavis-cummings model in circuit qed}. \newblock \emph{\bibinfo{journal}{Phys. Rev. Lett.}}   \textbf{\bibinfo{volume}{103}}, \bibinfo{pages}{083601}   (\bibinfo{year}{2009}). \bibitem{Simmonds_04} \bibinfo{author}{Simmonds, R.~W.} \emph{et~al.} \newblock \bibinfo{title}{Decoherence in {J}osephson phase qubits from junction   resonators}. \newblock \emph{\bibinfo{journal}{Phys. Rev. Lett.}}   \textbf{\bibinfo{volume}{93}}, \bibinfo{pages}{077003}   (\bibinfo{year}{2004}). \bibitem{TavisCummings_68} \bibinfo{author}{Tavis, M.} \& \bibinfo{author}{Cummings, F.~W.} \newblock \bibinfo{title}{Exact solution for an n-molecule-radiation-field   hamiltonian}. \newblock \emph{\bibinfo{journal}{Phys. Rev.}} \textbf{\bibinfo{volume}{170}},   \bibinfo{pages}{379--384} (\bibinfo{year}{1968}). \bibitem{QIP_09} \bibinfo{author}{Simmonds, R.~W.} \emph{et~al.} \newblock \bibinfo{title}{Coherent interactions between phase qubits, cavities,   and tls defects}. \newblock \emph{\bibinfo{journal}{Quantum Information Processing}}   \textbf{\bibinfo{volume}{8}}, \bibinfo{pages}{117--131}   (\bibinfo{year}{2009}). \bibitem{MartinisPRL05} \bibinfo{author}{Martinis, J.~M.} \emph{et~al.} \newblock \bibinfo{title}{Decoherence in {J}osephson qubits from dielectric   loss}. \newblock \emph{\bibinfo{journal}{Phys. Rev. Lett.}}   \textbf{\bibinfo{volume}{95}}, \bibinfo{pages}{210503}   (\bibinfo{year}{2005}). \bibitem{Cooper_04} \bibinfo{author}{Cooper, K.~B.} \emph{et~al.} \newblock \bibinfo{title}{Observation of quantum oscillations between a   {J}osephson phase qubit and a microscopic resonator using fast readout}. \newblock \emph{\bibinfo{journal}{Phys. Rev. Lett.}}   \textbf{\bibinfo{volume}{93}}, \bibinfo{pages}{180401}   (\bibinfo{year}{2004}). \bibitem{Hofheinz_09} \bibinfo{author}{Hofheinz, M.} \emph{et~al.} \newblock \bibinfo{title}{Synthesizing arbitrary quantum states in a   superconducting resonator}. \newblock \emph{\bibinfo{journal}{Nature}} \textbf{\bibinfo{volume}{459}},   \bibinfo{pages}{546--549} (\bibinfo{year}{2009}). \end{thebibliography}
\end{document}